# Spontaneous helix formation in polar smectic phase

Ewa Gorecka[a], Magdalena Majewska[a], Ladislav Fekete[b], Jakub Karcz[c] Julia Żukowska[c], Jakub Herman[c] Przemysław Kula[c], Damian Pociecha[a]

[a] *Faculty of Chemistry, University of Warsaw, Warsaw, Poland*
[b] *Institute of Physics, Academy of Sciences of Czech Republic, Prague, Poland*
[c] *Faculty of Advanced Technology and Chemistry, Military University of Technology, Warsaw, Poland*

**Abstract:** In soft ferroelectric crystals, the depolarization field can be reduced by periodic distortion of the polarization direction. In the polar nematic and tilted smectic phases, this process is energetically favoured , as it only requires changes in the director orientation. We demonstrate the spontaneous formation of a helical structure in the proper ferroelectric tilted smectic ($SmC_{TBF}$) phase, the phase is formed below the heliconical polar nematic ($N_{TBF}$) phase. The helical pitch in the smectic phase is approximately 600 nm and remains nearly constant across the entire temperature range of the phase. Under weak electric fields, the helix reorients while its structure remains largely intact; however, in stronger fields, the helix is destroyed as the electric polarization aligns along the electric field.

## Introduction

As suggested by the prefix 'ferro', ferroelectrics are analogous to ferromagnets. They belong to a broad class of materials characterized by a long-range order of electric dipoles, which results in spontaneous electric polarization, and they are distinguished by reversible switching of polarization with the application of an external electric field. Dipole order leads to uncompensated bound charges at the surfaces of ferroelectric films that generate an internal electric field, known as the depolarization field, which opposes the polarization. For a given electric polarization of 1 μC/cm², the internal electric fields within the crystal can reach magnitudes of up to $10^9$ V/m. In order to reduce the depolarization field, a ferroelectric crystal splits into domains, i.e. regions with different directions of polarization. In crystals, these domains are formed along specific crystallographic directions. The process of domain formation is driven by a balance of free energy: while the formation of domains reduces electrostatic energy, there is an energy penalty for the creation of domain walls [1]. In the recently discovered ferroelectric phase of soft matter - ferroelectric nematic phase, $N_F$, the depolarization energy can be reduced through different mechanisms. These include the formation of domains with splay deformation [2,3] or the formation of a helical structure [4]. Such mechanisms would be quite unusual for solid crystals, because it would require a periodic distortion of the crystal lattice [5,6]. However, in soft matter systems, like liquid crystals, in particular in polar nematics [7-10] that do not have periodic positional structure, this process is much easier as it requires only changes in director orientation. In liquid crystals, where the polarization is aligned along the director, the formation of a helix could fully cancel the polarization provided that the helical axis is perpendicular to the director as in cholesteric phase. This structure has been proposed theoretically as a ground state for $N_F$ phase [11, 12]. However the experimental evidence supporting this hypothesis is currently limited [13]. Alternatively, the polarization can be only partially compensated in the case of a heliconical structure, in which the director makes an oblique angle with respect to the helix axis. The heliconical structure has been confirmed for a variant of ferroelectric nematic phase (the phase was given the acronym $N_{TBF}$ to show that the structure is similar to the twist bend nematic phase; however, experiments show much longer helix), and a distinct phase transition has been observed between the uniform $N_F$ and the helical $N_{TBF}$ phases [4]. The heliconical structure was also suggested for the axially polar tilted smectic phase [4, 14]. In this communication we demonstrate convincing evidence for the sequence of ferroelectric nematic and ferroelectric smectic phases, both of which exhibit spontaneous helix formation.

## Results and Discussion

The studied material (see Fig. 1a for its molecular structure and the Supporting Information for details of its synthesis) exhibits a sequence of five liquid crystalline phases upon cooling from the isotropic liquid phase: Iso → N → $N_x$ → $N_F$ → $N_{TBF}$ → $SmC_{TBF}$. Upon rapid cooling recrystallization of the material could be easily prevented, and the smectic phase observed at room temperature. This phase sequence was confirmed by differential scanning calorimetry (DSC) measurements (Figure S1, Table S1).It was observed that each phase transition was accompanied by a distinct thermal effect, with the exception of the $N_F$-$N_{TBF}$ phase transition where enthalpy changes were below the detection limit (<0.001 J/g). The subsequent LC phases are as follows: paraelectric N, antiferroelectric $N_X$ (sometimes named also $SmZ_A$ [15,16]) with a periodic structure of domains with antiparallel direction of polarization, ferroelectric $N_F$, $N_{TBF}$ phase – a ferroelectric phase with a heliconical structure and the heliconical tilted smectic $SmC_{TBF}$. The X-ray diffraction studies (Figure 1) further validated the sequence of nematics-smectic phase: all four nematic phases display only diffuse diffraction signals characteristic of short-range positional order. However, the low-angle signal, related to the longitudinal distance between molecules, gradually narrows as the temperature decreases, indicating increasing positional correlations in the material. In the smectic phase, this signal becomes limited by the machine's resolution, proving true long-range order of molecular positions. Furthermore, the signal moves to smaller angles which indicates a reduction in layer spacing, as the tilt increases with decreasing temperature. Conversly, the high-angle diffraction signal, associated with the transverse distance between molecules remains diffuse across all the LC phases, which is characteristic for a liquid-like order.

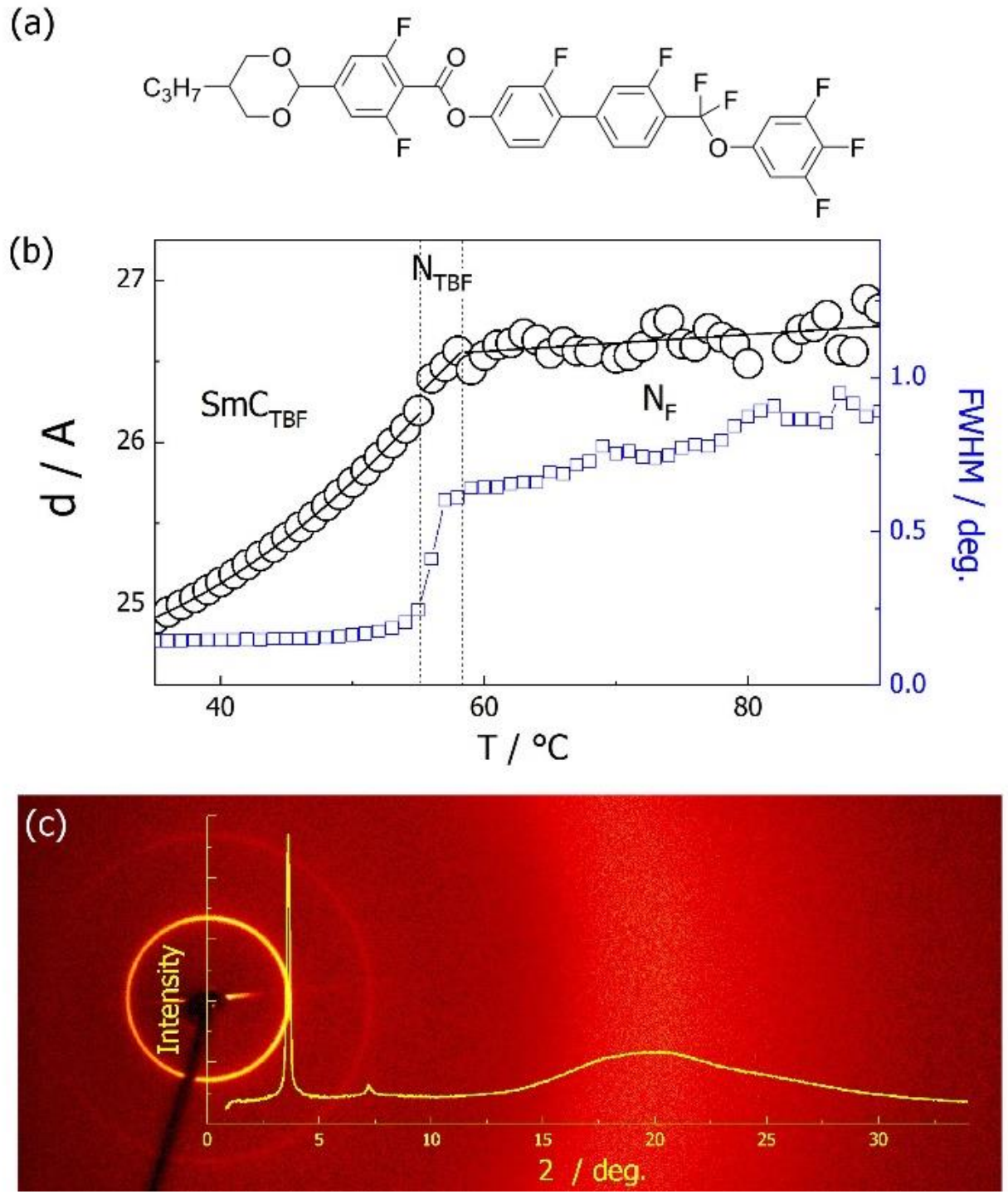

**Figure 1.** (a) Molecular structure of the studied compound. (b) Layer spacing, d, (in $SmC_{TBF}$ phase) and averaged longitudinal distance of molecules in $N_F$ and $N_{TBF}$ phases deduced form X-ray diffraction studies (black circles) and the width (FWHM) of the related diffraction signal (blue squares) measured vs. temperature; note that the signal becomes instrumental-resolution-limited in smectic phase. (c) 2D X-ray diffraction pattern recorded in $SmC_{TBF}$ phase with superimposed intensity vs. diffraction angle obtained by integration of the pattern over azimuthal angle. The high angle signal is diffused, proving liquid like order of molecules inside the smectic layers.

The orientational order of molecules was assessed by optical birefringence (Δn) measurements (Figure 2), performed in cells, having planar aligning polymer layers rubbed parallel to secure the same anchoring conditions at both surfaces (this allows to avoid formation of twisted domains along cell thickness). In the N and $N_x$ phases the birefringence monotonically increases on cooling, however with slightly different slope, following growth of orientational order of molecules. A small jump at the transition to the $N_F$ phase indicates that appearance of polar order is accompanied by weak, step-like increase in orientational order. In the $N_{TBF}$ phase Δn decreases, as the tilting of molecules and formation of a helical structure reduces the refractive index along the helix axis and increases refractive index perpendicular to the helix [17]. The continuous changes of birefringence indicate that at the $N_F$-$N_{TBF}$ phase transition tilt increases continuously from zero and reaches ~25 deg. before the transition to the $SmC_{TBF}$ phase. The decrease of birefringence accompanied by the increase of molecular tilt continues in the smectic phase, however the values of Δn in the SmC phase are not fully reliable, because the sample alignment becomes worse in this phase.

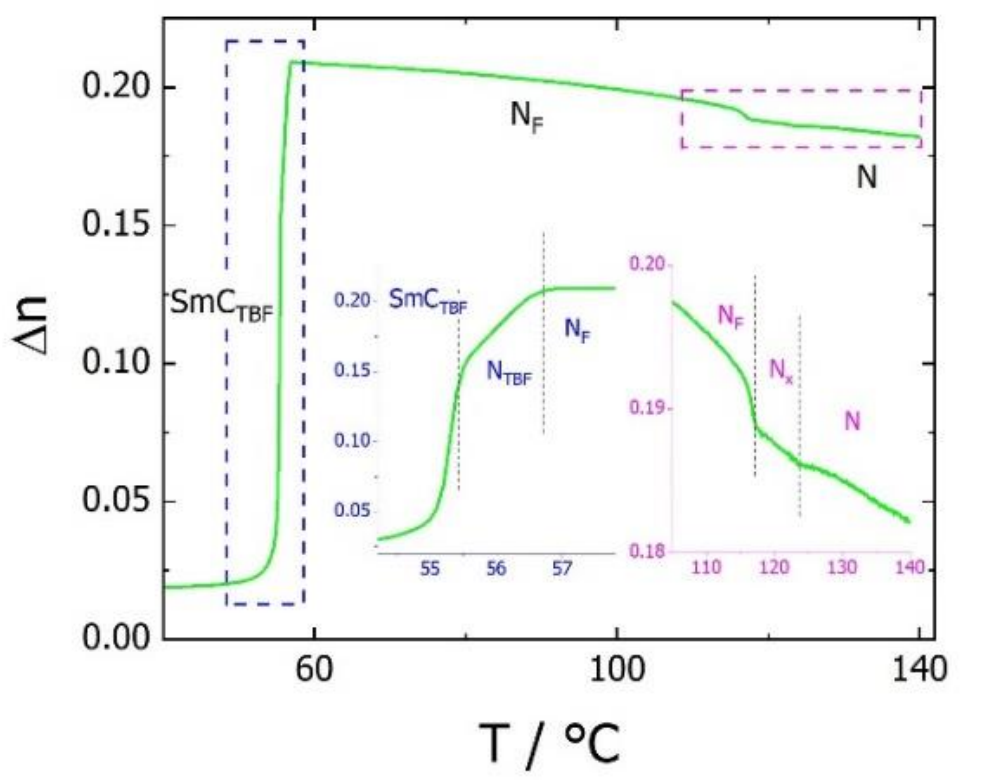

**Figure 2.** Optical birefringence vs. temperature measured in thin (~1.8-μm-thick) planar cell with parallel rubbing at both surfaces.

The polar properties of the nematic and smectic phases were confirmed by dielectric spectroscopy and polarization reversal current measurements. Although interpretation of dielectric spectroscopy results for strongly polar materials can be challenging [18, 19], the data clearly indicate, that upon entering the $N_F$ phase the dipole order becomes long-range and a strong dielectric response emerges (Figure S2). Above this temperature in the paraelectric N and antiferroelectric $N_x$ phases, the dielectric response is weak and no distinct dielectric modes are observed in the studied frequency range. In the $N_F$ and $N_{TBF}$ phases, a strong dielectric response is observed at frequencies below 10 kHz, the mode weakens and shifts to lower frequencies in the smectic phase. Polarization switching current peak was detected under application of triangular voltage in the $N_F$, $N_{TBF}$, and $SmC_{TBF}$ phases (Figure S3). Far from the phase transition to the polar phase, the spontaneous electric polarization is ~4.5 μC $cm^{-2}$, which is a typical value for this class of materials with nearly perfect order of all molecular dipoles [20]. In the smectic phase the threshold voltage, at which the polarization reversal takes place starts to increase on cooling.

When observed under the polarized light microscope in a cell with polymer anchoring layer with parallel rubbing on both surfaces, the texture of the N phase is uniform, with the extinction directions along the polarizers. In the $N_F$ phase the texture is essentially unchanged with few defects developing around the glass spacers (Figure 3), in which the director departs in a parabolic way from rubbing direction [21]. In the $N_{TBF}$ phase the texture becomes more complex, with stripes parallel to the rubbing direction, the period of these stripes is dependent on the cell thickness. In the $SmC_{TBF}$ phase the texture is similar, however the stripes are less regular. The light diffraction experiment revealed that in addition to the aforementioned stripes, which are easily detectable in microscopic observations, the sample has also additional, sub-micron periodicity in both $N_{TBF}$ and $SmC_F$ phases, due to the helical pitch. It was observed that in the $N_{TBF}$ phase the helix unwinds on heating as the $N_F$ phase is approached (Figure 4), and this allowed to directly observe related periodicity by microscopy, as a regular array of strips with ~1 μm spacing, running perpendicular to the cell rubbing direction (Figure 3).

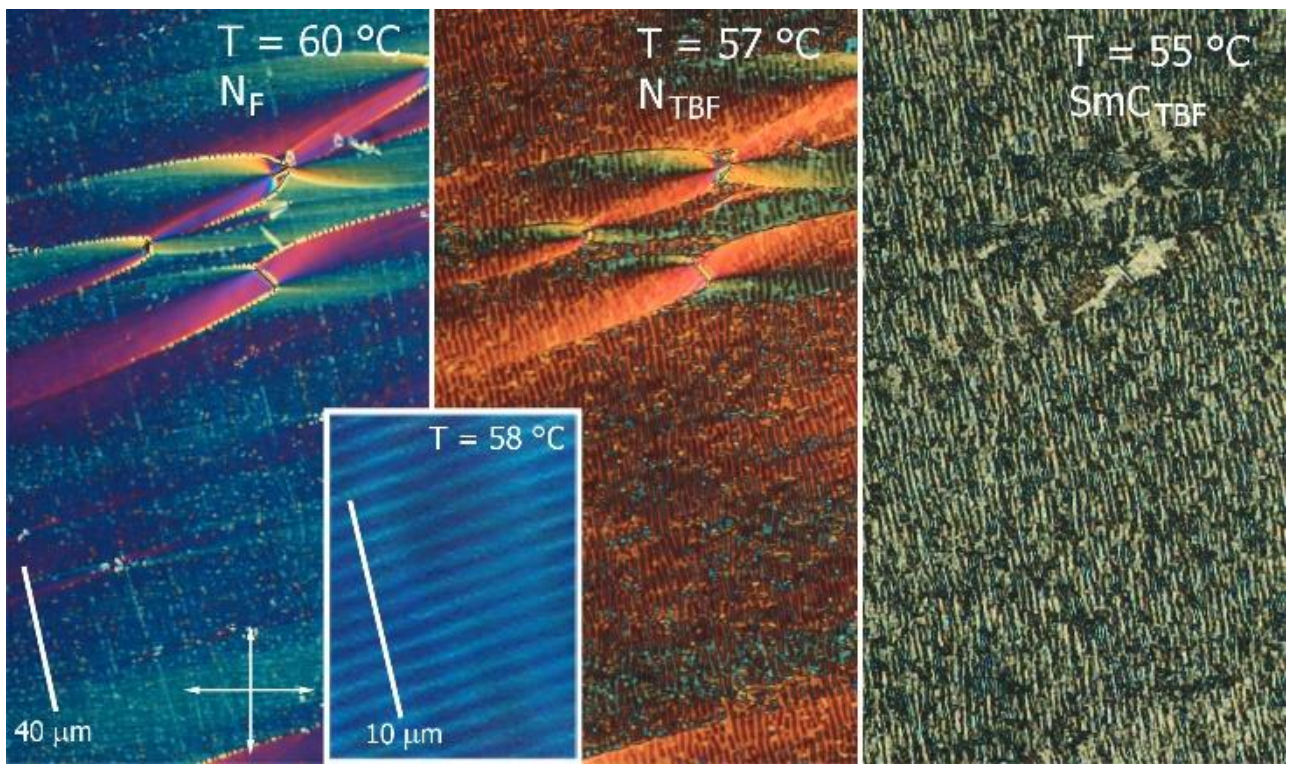

**Figure 3.** Textures of $N_F$, $N_{TBF}$ and $SmC_{TBF}$ phase in planar cell with parallel rubbing direction, stripes visible in the image of $N_{TBF}$ and $SmC_{TBF}$ are related to the cell thickness, the stripes related to the helix (in the inset) can be observed only in the vicinity of $N_{TBF}$-$N_F$ phase transition where the helix unwinds.

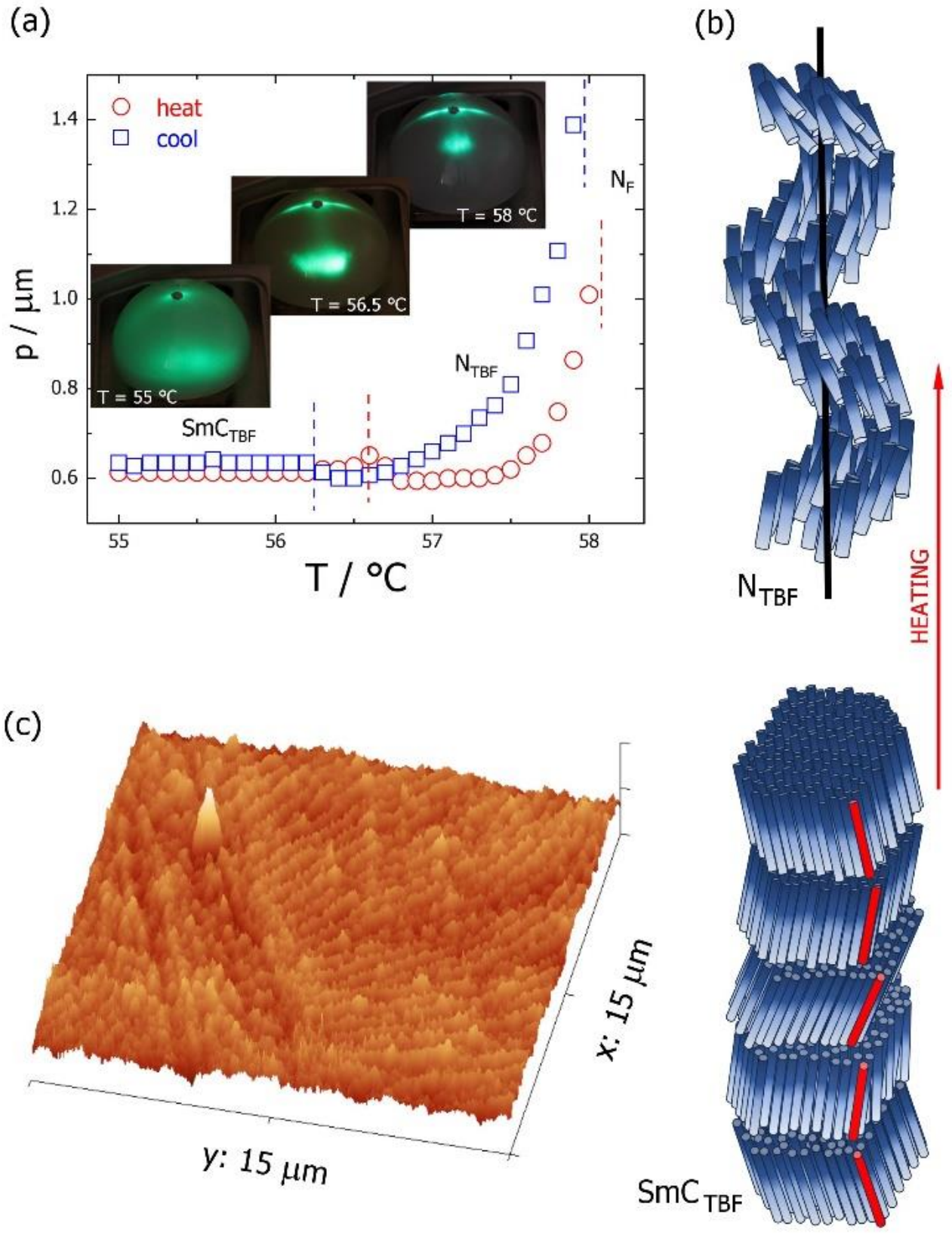

**Figure 4.** (a) Helical pitch vs. temperature and in the inset light diffraction patterns in $N_{TBF}$ and $SmC_{TBF}$ phases, the intensive spots are from diffraction on helical structure. (b) Models of heliconical arrangement of molecules in $N_{TBF}$ and $SmC_{TBF}$ phases, color gradient of rods reflects dipolar character of molecules. (c) AFM image of the $SmC_{TBF}$ phase supercooled to room temperature, the stripe periodicity is ~600-700 nm.

Near the transition to the $SmC_{TBF}$ phase, the helical pitch of the $N_{TBF}$ phase reaches around 600 nm and remains almost constant in the $SmC_{TBF}$ phase. The material can be easily supercooled to room temperature, allowing the confirmation of the periodic structure by atomic force microscopy (AFM) (at higher temperatures, adhesion is too strong and viscosity too low for such measurements). A clear pattern with a periodicity of 600-700 nm was observed, consistent with light diffraction measurements (Figure 4c, Figure S4). The piezoresponse force microscopy (PFM) measurements confirmed that in the area exposed to a strong electric field, electric polarization becomes re-oriented along the field direction and the orientation of polarization remains after the field is removed (Figure S5).

The $SmC_{TBF}$ phase was also studied under applied electric field, two different sample geometries were used: (*i*) planar cells with parallel rubbing at both surfaces and electric field applied across the cell thickness, and (*ii*) planar cells with parallel rubbing at both surfaces and in-plane electric field perpendicular to the rubbing direction. In the first case weak electric field caused re-arrangement of the smectic layers from planar (bookshelf) to homeotropic alignment (with helix axis along the electric field). The field-treated sample exhibited significantly lower birefringence compared to the untreated texture and displayed several tens of micrometer-size domains with optical activity (Figure 5). These domains had a randomly distributed sign of optical activity, demonstrating that both left- and right-handed helices are formed with equal probability. Applying a stronger electric field disrupted the helices by aligning the dipole moments within the layers along the direction of the field, which lead to the disappearance of optical activity of the texture.

Reorientation of the helix axis direction in the cells with an in-plane electric field was monitored by observing laser diffraction patterns. As expected, prior to the application of electric field, the diffraction spots, related to the helical structure, were positioned along the rubbing direction. Above a certain threshold, the electric field caused a rotation of the helix axis, as evidenced by the appearance of the diffraction spots positioned along the electric field (perpendicular to the rubbing direction). However, reducing the field to zero did not fully restore the original helix orientation; instead, two sets of diffraction spots were visible, originating from distinct areas in the sample with the helix oriented in perpendicular directions (Figure 6a).

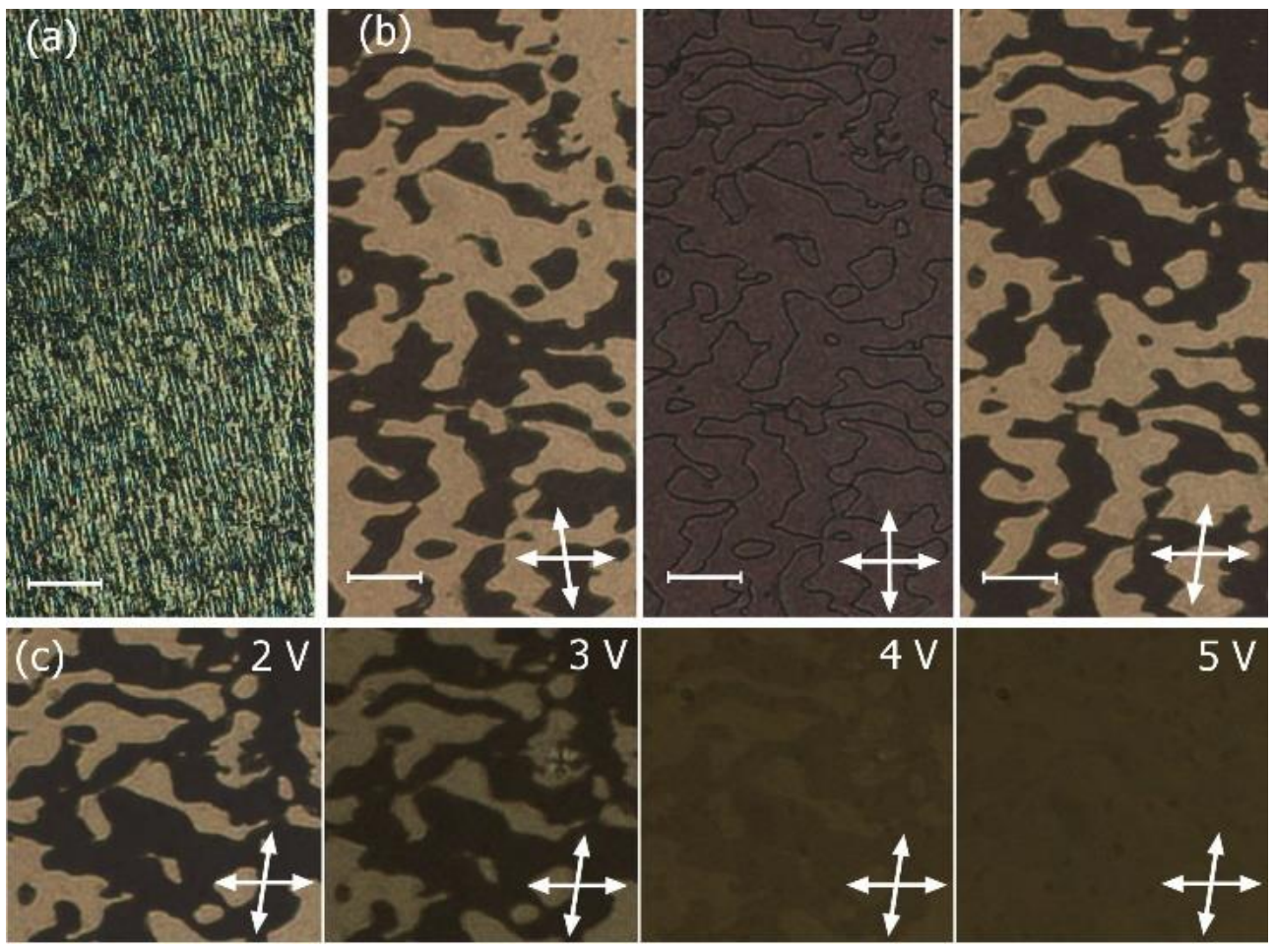

**Figure 5.** (a) Texture of $SmC_{TBF}$ phase obtained upon cooling the sample from $N_{TBF}$ phase in 3-μm-thick cell with parallel rubbing on both surfaces. (b) Application of weak electric field across the cell thickness (<1V/μm), realigned the helix form planar to homeotropic position, this orientation of helix remains after removing electric field. By uncrossing the polarizers (±10 degree) optically active domains are detected showing that the left- and right-handed helices coexist. (c) Under increasing electric field applied across the cell thickness, optical activity diminishes, in sufficiently strong electric field the helix is destroyed and the uniform state with electric polarization aligned along the electric field is obtained (homeotropic texture).

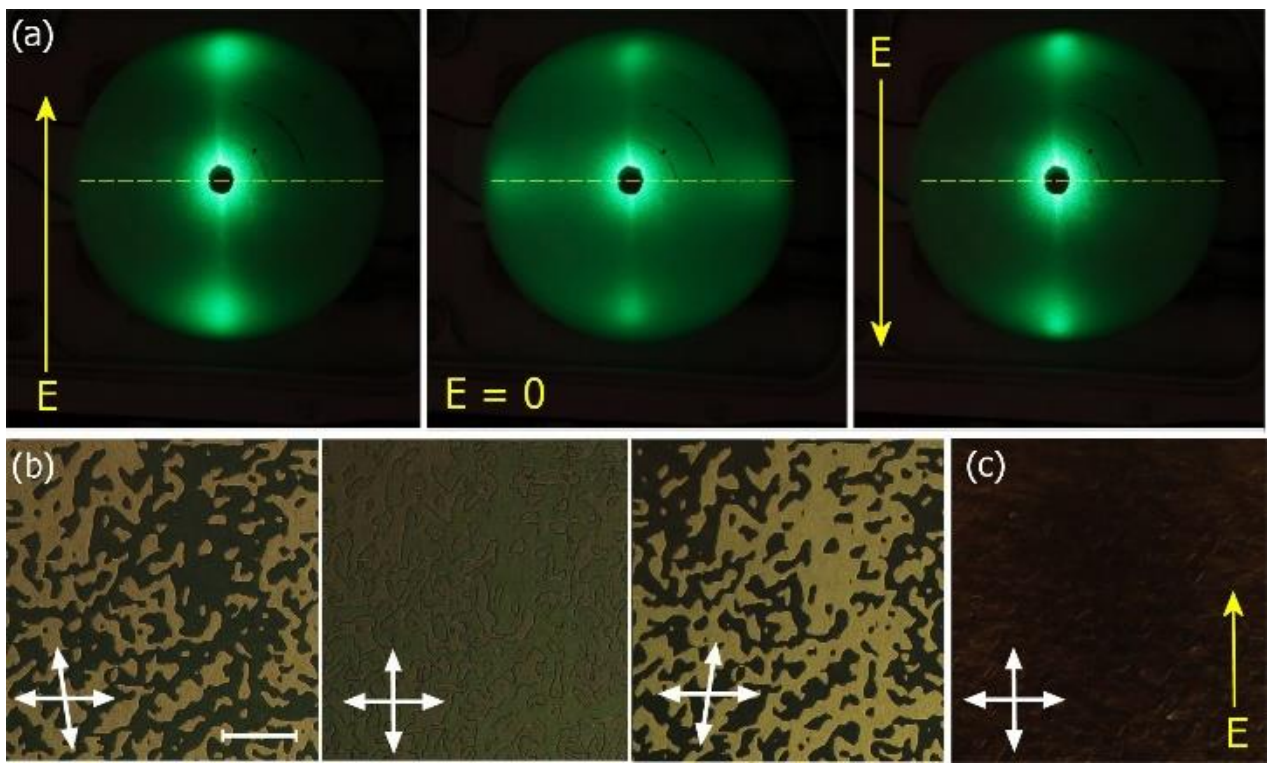

**Figure 6.** Laser light diffraction patterns in $SmC_{TBF}$ phase from the planar cell ~ 2.5-μm-thick with in-plane electrodes. (a) Under applied voltage only the diffraction spots along the field directions are visible. When electric field is switched off two sets of diffraction spots are visible, along and perpendicular to the rubbing direction (yellow dashed lines). (b) optical texture in the same cell observed after application of a.c. electric field, domains with opposite sign of optical activity are clearly visible by slight de-crossing of polarizers (arrows). (c) The same sample area under applied in-plane electric field, E=50 V mm$^{-1}$. The optical axis (layer normal) is along the electric field.

Apparently, field treated sample is divided along its thickness, in the bulk of the sample helix remains aligned along the direction in which the electric field had been applied, while in the surface regions the helix is oriented along the rubbing direction. The application of an a.c. electric field results in sequential changes of the diffraction spot intensities, as the relative size of the bulk and surface regions changes (movie S1). Although in both, bulk and surface regions, helix axis is still oriented perpendicular to the sample surface, after switching-off the field the sample exhibited texture with optically active domains (Figure 6b). This suggest, that those regions are interconnected by screw dislocations, such a boundary facilitates easy growth of one region at the expense of the other under electric field. The process of the reorientation of the helical axis direction and the distortion of the helix has been also followed by monitoring the changes of the azimuthal direction of the optical axis and birefringence (Figure S6). Two clear regimes were detected, at weak electric field mainly changes of the optical axis direction were found, while above certain threshold strong increase in the birefringence was observed due to a decrease of the tilt angle of the heliconical structure.

## Conclusion

Recently unambiguous evidences were given, that apart from the polar, orthogonal $SmA_F$ phase with uniform electric polarization along the director also its tilted version - the polar smectic $C_F$ phase exists [22, 23]. In these phases, similarly as in the polar nematic $N_F$ phase, a strong internal electric field (depolarization field) might be reduced the formation of random structures of polar domains. Here we show the alternative pathway, wherein a strong depolarization field is reduced by spontaneously formed helical structure. This structure is a smectic analogue of recently discovered $N_{TBF}$ phase. In the tilted smectic phase, the helical pitch is in sub-micron range and the transition from $N_{TBF}$ to $SmC_{TBF}$ phase occurs without significant changes in helical pitch. The axis of the heliconical structure can be reoriented by a small electric field while its structure remains intact.

**Experimental methods:**

***Calorimetric studies:*** For differential scanning calorimetry (DSC) studies, a TA Q200 calorimeter was used, calibrated using indium and zinc standards. Heating and cooling rates were 5-20 K $min^{-1}$, samples were kept in a nitrogen atmosphere. The transition temperatures and associated thermal effects were extracted from the heating traces.

***X-ray diffraction:*** X-ray diffraction (XRD) studies in broad diffraction angle range were performed with Bruker GADDS system equipped with micro-focus type X-ray tube with Cu anode, and Vantec 2000 area detector. Samples were prepared in the form of small drops placed on a heated surface, their temperature was controlled with a modified Linkam heating stage. For small angle diffraction experiments Bruker Nanostar system was used (micro-focus type X-ray tube with Cu anode, MRI TCPU-H heating stage, Vantec 2000 area detector). Samples were prepared in thin-walled glass capillaries, with 1.5 mm diameter.

**Microscopic Studies:** Optical textures of LC phases were studied using a Zeiss Axio Imager A2m polarized light microscope, equipped with a Linkam TMS 92 heating stage. Samples were prepared in commercial cells (AWAT) of various thicknesses (1.5–20 μm) with ITO electrodes and surfactant layers for planar or homeotropic alignment, in the case of planar cells either parallel or antiparallel rubbing on both surfaces was applied. Cells for in-plane switching were provided by prof. O. Lavrentovich group at Kent State University.

***Optical birefringence:*** Optical Birefringence was measured with a setup based on a photoelastic modulator (PEM-90, Hinds) working at the base frequency f=50 kHz. As a light source, a halogen lamp (Hamamatsu LC8) equipped with a narrow bandpass filter (532±3 nm) was used. Samples were prepared in glass cells with a thickness of 1.5 μm, having surfactant layers for planar anchoring condition, and parallel rubbing assuring uniform alignment of the optical axis in nematic phases. The sample and PEM were placed between crossed linear polarizers, with axes rotated ±45 deg with respect to the PEM axis, and the intensity of the light transmitted through this set-up was measured with a photodiode (FLC Electronics PIN-20). The registered signal was de-convoluted with a lock-in amplifier (EG&G 7265) into 1f and 2f components to yield a retardation induced by the sample. Based on the measured optical birefringence the conical tilt angle ($\theta$) in the twist-bend ferroelectric nematic phase ($N_{TBF}$) was deduced from the decrease of the $\Delta n$ with respect to the values measured in the ferroelectric nematic ($N_F$) phase, according to the relation: $\Delta n_{NTBF} = \Delta n_{NF} (3\cos^2\theta - 1)/2$ [17]. The birefringence of the ferroelectric nematic phase was extrapolated to the lower temperature range by assuming a power law temperature dependence: $\Delta n_{NF} = \Delta n_0\,(Tc - T)^{\gamma}$ , where $\Delta n_0$, $Tc$ , and $\gamma$ are the fitting parameters.

***Laser diffraction studies:*** Optical diffraction studies were performed for the samples placed on a heating stage and illuminated from below with green (520 nm) laser light. The diffraction pattern was recorded on the half-sphere screen placed above the sample, the angular position of the observed diffraction signal allowed for calculation of the related periodicity of the stripe pattern in the cell, and helical pitch length.

***Dielectric spectroscopy:*** The complex dielectric permittivity was measured in the 1 Hz–10 MHz frequency range using a Solartron 1260 impedance analyzer. The material was placed in 3-μmthick glass cell with ITO electrodes (without the polymer alignment layers to avoid the influence of the high capacitance of a thin polymer layer). The amplitude of the applied ac voltage, 20 mV, was low enough to avoid Fréedericksz transition in nematic phases.

***Atomic Force microscopy (AFM) and Piezoresponse Force Microscopy (PFM):*** AFM measurements were performed using a Bruker Dimension Icon Microscope working in tapping or Peak Force Quantitative Nanomechanics (QNM) mode and cantilevers with 0.4 N/m force constant were applied. The PFM experiments were conducted with the same instrument, in an optimized vertical domains mode. In this setup, the deflection of a conducting AFM cantilever is detected as the piezoelectric material's domains deform mechanically in response to an applied voltage. Due to the small magnitude of these displacements, a lock-in technique is used. A modulated reference voltage is applied to the AFM tip, inducing deformation of the sample surface. The AFM tip remains in contact mode during the measurement. The lock-in amplifier detects the cantilever's deflection signal, which oscillates in phase with the reference drive signal during deformation, and out of phase (perpendicular) when no deformation occurs. These signals are then converted into amplitude and phase angle data for the cantilever deflection. Ideally, at the domain wall, the amplitude should drop to zero, and the phase should shift by 180 degrees.

The sample was placed on an ITO electrode surface, which was electrically connected to the AFM stage. The PFM measurements were performed with dedicated to soft material use Budget Sensors AIOE cantilevers with a resonance frequency of 15 kHz and a 0.2 N/m force constant.

**Organic synthesis and analytical data:**

Studied here material, 4'-(difluoro(3,4,5-trifluorophenoxy)methyl)-2,3'-difluoro-[1,1'-biphenyl]-4-yl 2,6-difluoro-4-(5-propyl-1,3-dioxan-2-yl)benzoate was synthesized at Liquid Crystal group of the Military University of Technology is modification of JK103 structure [4]. In first step , 2,6-difluoro-4-(5-propyl-1,3-dioxan-2-yl)benzoic acid **(3)** was synthesized following the procedure in [24].

In order to purify the **(3)** to a pure *trans* form of 1,3-dioxane unit, a series of recrystallization processes were performed, after which the compound was constantly enriched with the *trans* isomer. The recrystallisation process was repeated until > 99% isomeric purity was achieved.

Next, during the Suzuki-Miyaura process the phenol derivative **(3)** was obtained by reacting boronic pinacol ester derivative **(2)**, obtained previously from bromo derivative **(1)** in a Miyaura borylation reaction, with the 4-bromo-3-fluorophenol. Final compound was synthesized in a Steglich esterification conditions between **(3)** and **(4)** with the presence of DCC and DMAP. The compound was purified using sequence of recrystallization and liquid column chromatography techniques.

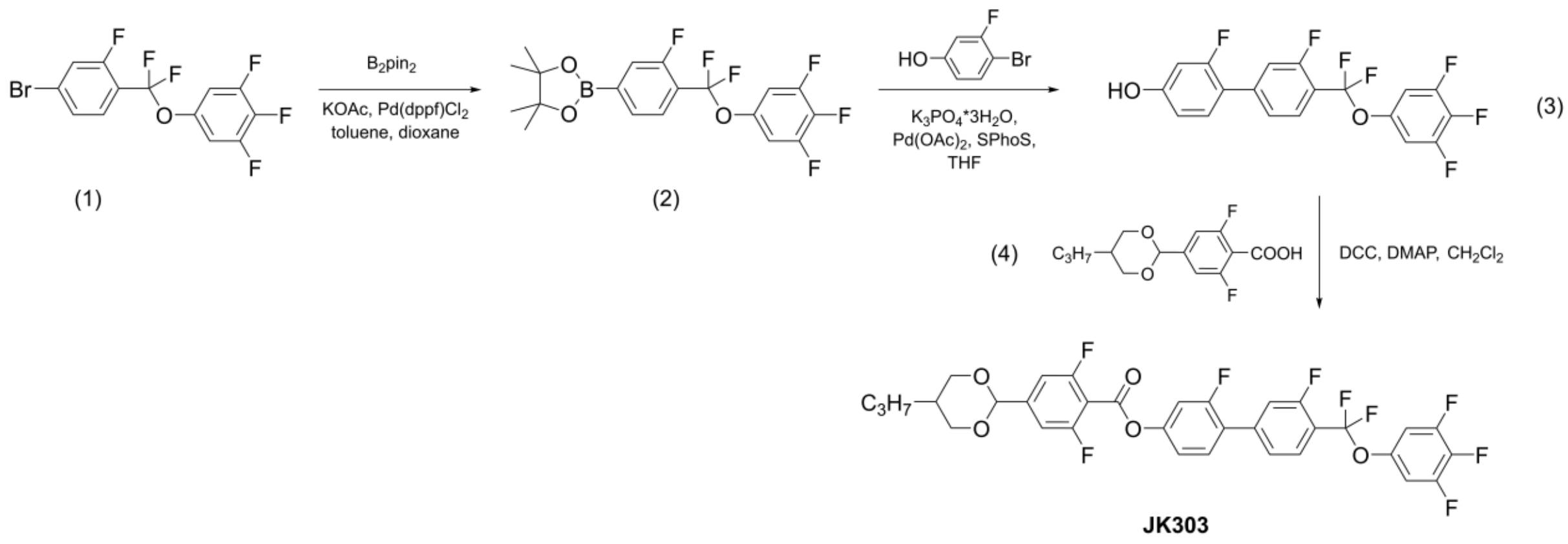

**Fig. S1.** Synthesis of studied compound 4'-(difluoro(3,4,5-trifluorophenoxy)methyl)-2,3'-difluoro-[1,1'-biphenyl]-4-yl 2,6-difluoro-4-(5-propyl-1,3-dioxan-2-yl)benzoate

***2-(4-(difluoro(3,4,5-trifluorophenoxy)methyl)-3-fluorophenyl)-4,4,5,5-tetramethyl-1,3,2,-dioxaborolane (2)***

5-((4-bromo-2-fluorophenyl)difluoromethoxy)-1, 2,3-trifluorobenzene **(1)** (6.6 g - 0.018 mol), bis(pinacolato)diboron (4.8 g - 0.019 mol), potassium acetate (5.29 g - 0.054 mol), toluene and dioxane (1:1 vol/vol) were placed in flask. The reaction was refluxed under a nitrogen atmosphere. $PdCl_2$(dppf) (0.4 g - 0.00054 mol) was then added. The reaction was refluxed for 48 h under a nitrogen atmosphere. The reaction was poured into water and filtered under reduced pressure through a Fuller's earth plate. The phases were separated by washing the organic phase with water and the aqueous phase with toluene. The organic phase was dried over anhydrous $MgSO_4$ and concentrated. Product was purified using chromatography column (silica gel, hexane).

Yield 5.2 g (70%).
Purity 95% (GC-MS)
m.p. = 55-57°C
MS(EI) m/z: 418 (M+); 403; 319; 255; 241; 229; 213; 189; 171; 144; 131; 119

$^{1}$H NMR (500 MHz, $CDCl_3$) δ: 7.67 (m, 2 H); 7.61 (d, *J*=10.99 Hz, 1 H); 6.98 (t, J=6.85 Hz, 2H); 1.38 (s, 12 H)

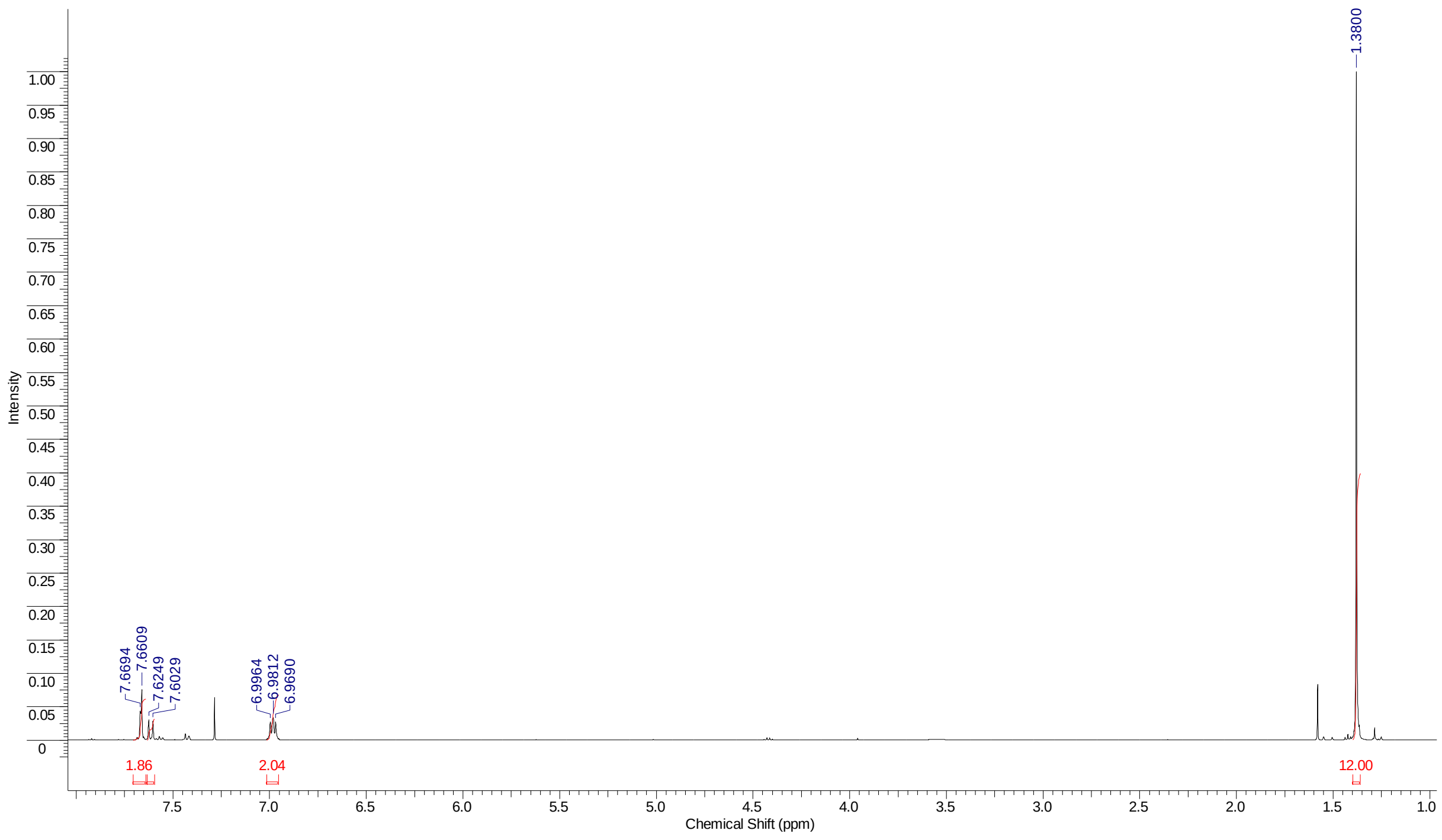

$^{13}$C NMR (125 MHz, $CDCl_3$) δ: 158.38; 152.00 (dd, *J*=10.90, 5.45 Hz); 150.00 (dd, J= 10.83, 5.45 Hz); 144.96 (m); 139.33 (t, *J*=14.99 Hz); 137.34 (t, *J*=15.44 Hz); 130.07 (d, *J*=3.63 Hz); 126.73 (t, *J*=5.00 Hz); 122.56 (m); 122.40; 120.79; 107.41 (m); 84.57; 24.87

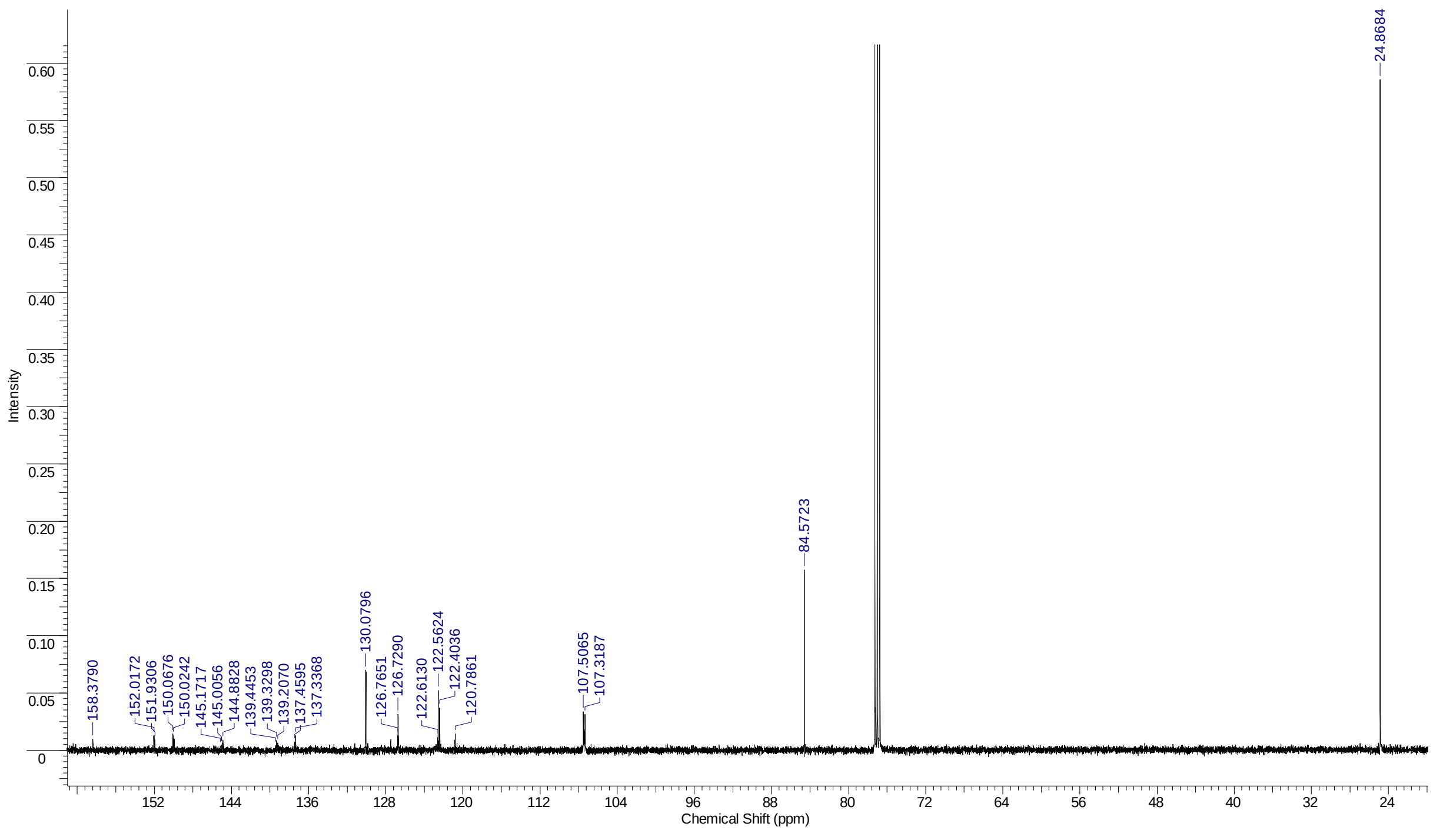

***4'-(difluoro(3,4,5-trifluorophenoxy)methyl)-2,3'-difluoro-[1,1'-biphenyl]-4-ol (3)***

2-(4-(difluoro(3,4,5-trifluorophenoxy)methyl)-3-fluorophenyl)-4,4,5, 5-tetramethyl-1,3,2,-dioxaborolate **(2)** (3.7 g - 0.009 mol), 4-bromo-3-fluorophenol (1.54 g - 0.008 mol), $K_3PO_4$ * $3H_2O$ (8.4 g - 0.031 mol) and 250 $cm^3$ of anhydrous tetrahydrofuran were placed in flask. The reaction was refluxed under a nitrogen atmosphere. $Pd(OAc)_2$ (0.047 g - 0.00022 mol) and SPhos (0.092 g - 0.00022 mol) were then added. The reaction was refluxed for 24 h under a nitrogen atmosphere. Reaction was washed with water and dichloromethane; filtered off; product was extracted with dichloromethane; the organic layer was washed with water, dried over anhydrous $MgSO_4$ and concentrated. Phenol was concentrated and purified using chromatography column (silica gel, dichloromethane).

Yield 1.5 g (47%).
Purity 92% (GC-MS)
m.p. = 94-96°C
MS(EI) m/z: 402 (M+); 383; 331; 297; 255; 235; 226; 206; 186; 175; 157; 147; 119

$^1$H NMR (500 MHz, $CDCl_3$) δ: 7.72 (t, *J*=7.93 Hz, 1 H); 7.36 (m, 3 H); 7.01 (dd, *J*=7.63, 6.10 Hz, 2 H); 6.74 (m, 2 H)

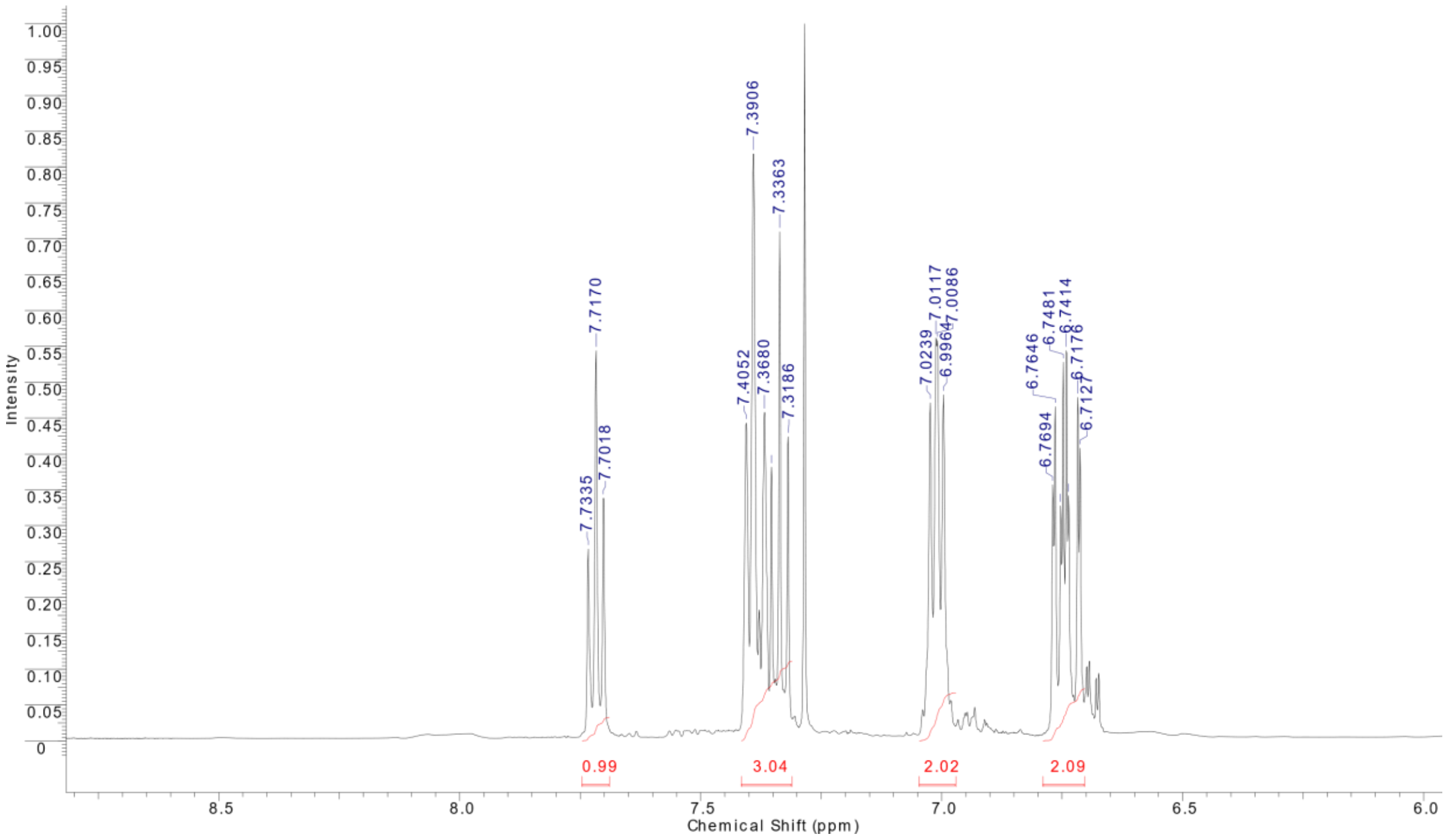

$^{13}$C NMR (125 MHz, $CDCl_3$) δ: 162.29; 159.31; 157.54 (d, *J*=11.8 Hz); 150.99 (dd, *J*=250.3, 16.3 Hz); 131.04; 127.30; 124.04; 118.96 (t, *J*=10.0 Hz); 116.98 (d, *J*=18.2 Hz); 107.51 (d, *J*=21.8 Hz); 104.02 (d, *J*=26.3 Hz)

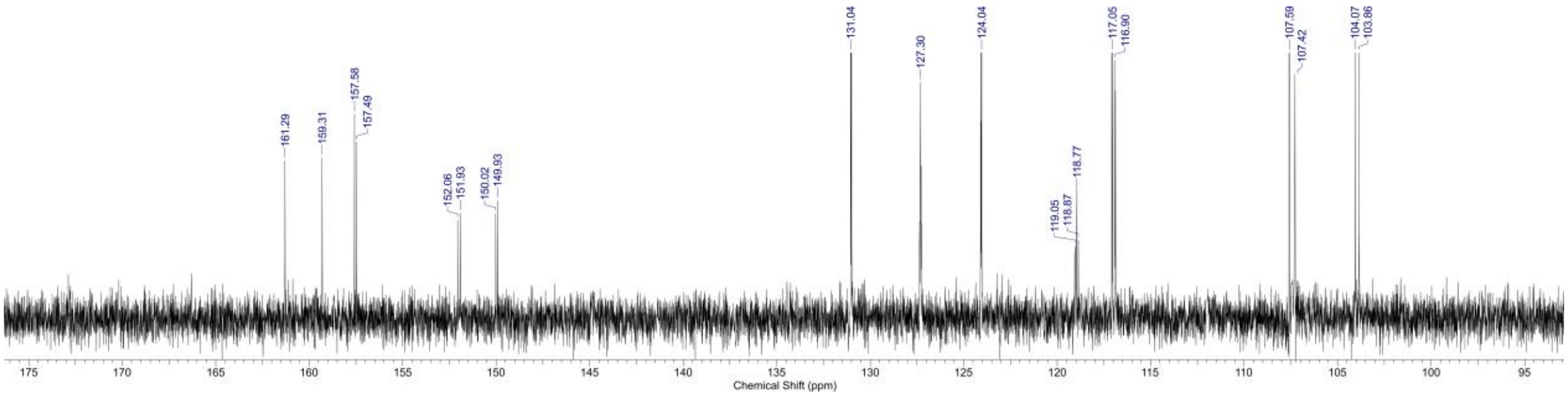

***4'-(difluoro(3,4,5-trifluorophenoxy)methyl)-2,3'-difluoro-[1,1'-biphenyl]-4-yl 2,6-difluoro-4-(5-propyl-1,3-dioxan-2-yl)benzoate***

To a stirred solution of 2,6-difluoro-4-(5-propyl-1,3-dioxan-2-yl)benzoic acid **(4)** (1.0 g; 0.003 mol), 4'-(difluoro(3,4,5-trifluorophenoxy)methyl)-2,3'-difluoro-[1,1'-biphenyl]-4-ol **(3)** (1.5 g; 0.0037 mol) and N,N'-dicyclohexylcarbodiimide DCC (0.8 g; 0.004 mol) in dichloromethane, DMAP (0.1g) was added and the solution was stirred overnight at room temperature. The reaction mixture was filtered through silica pad and the filtrate was concentrated under vacuum. The product was purified using sequence of recrystallization (ethanol/acetone mixture) and liquid column chromatography (silica gel and dichloromethane) techniques to give white solid.

Yield 0.9g (36%)
Purity 99.5% (99.6/0.2 trans/cis ratio) (HPLC-MS)
MS(EI) m/z: 670 (M+); 652; 523; 423; 401; 369; 269; 254; 226; 206; 185; 169; 141

$^{1}$H NMR (500 MHz, $CDCl_3$) δ: 7.77 (t, *J*=7.93 Hz, 1 H); 7.52 (t, *J*=8.70 Hz, 1 H); 7.44 (m, 2 H); 7.21 (m, 4 H); 7.02 (dd, *J*=7.63, 5.80 Hz, 2 H); 5.43 (s, 1 H); 4.28 (dd, *J*=11.75, 4.73 Hz, 2 H); 3.56 (t, *J*=11.44 Hz, 2 H); 2.16 (m, 1 H); 1.37 (m, 2 H); 1.13 (m, 2 H); 0.96 (t, *J*=7.32 Hz, 3 H)

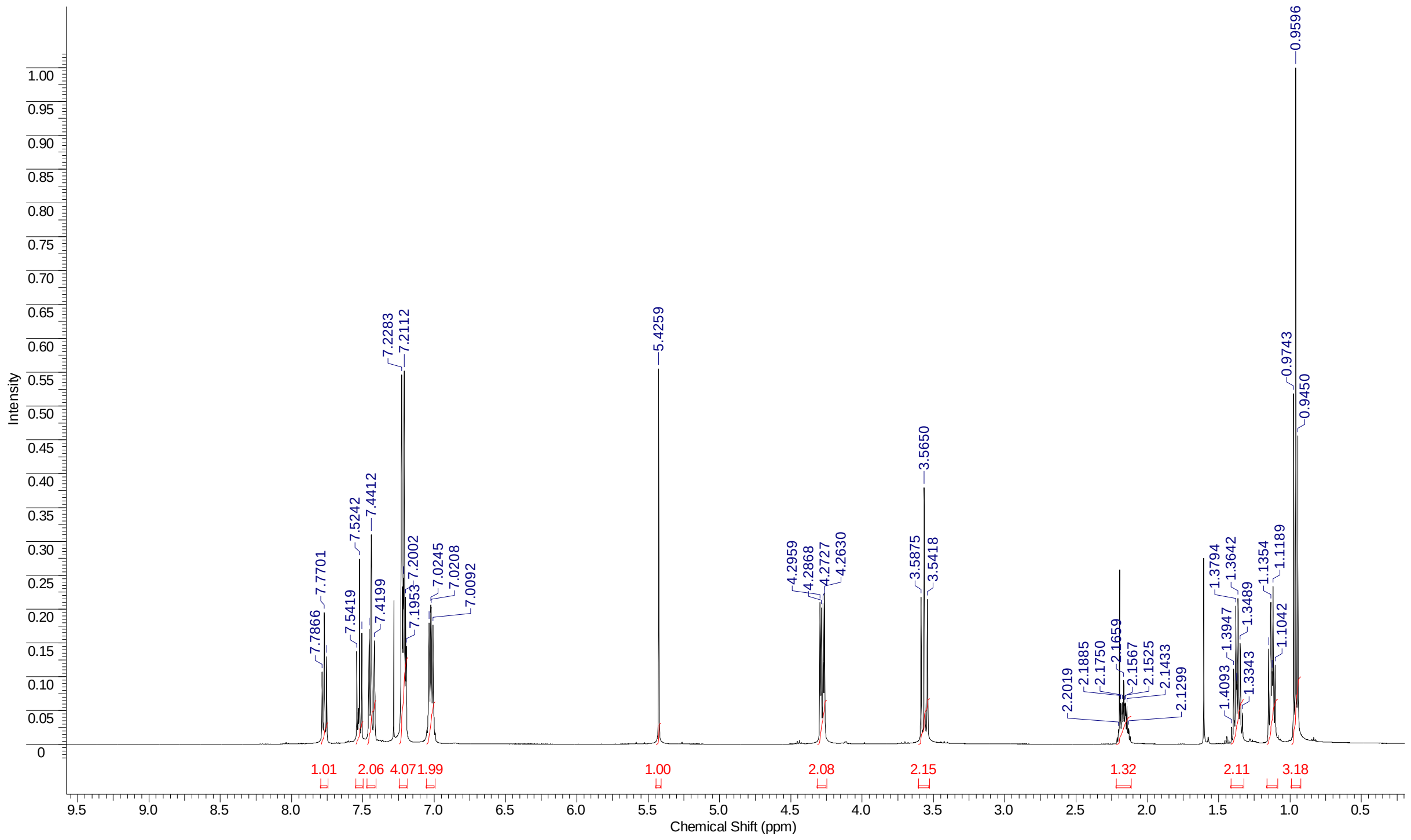

$^{13}$C NMR (125 MHz, $CDCl_3$) δ: 161.97 (d, *J*=5.45 Hz); 160.75; 160.55; 159.92 (d, *J*=5.45 Hz); 159.28; 158.71; 158.55; 152.04 (dd, *J*=10.90, 5.45 Hz); 151.23 (d, *J*=10.90 Hz); 150.04 (dd, *J*=10.90, 5.45 Hz); 145.68 (t, *J*=9.99 Hz); 144.87 (m); 140.58 (d, *J*=8.17 Hz); 139.37 (t, *J*=15.44 Hz); 137.39 (t, *J*=15.44 Hz); 130.87 (d, *J*=3.63 Hz); 127.54 (m); 124.75 (dd, *J*=13.62, 1.82 Hz); 124.47 (t, *J*=3.18 Hz); 120.74 (t, J=264,31 Hz); 119.73 (m); 118.16 (d, *J*=3.63 Hz); 117.41 (dd, *J*=22.71, 3.63 Hz); 110.66 (d, *J*=25.43 Hz); 110.32 (dd, *J*=23.16, 3.18 Hz); 109.46 (t, *J*=17.26 Hz); 107.43 (m); 98.82; 72.60; 33.92; 30.25; 19.55; 14.18

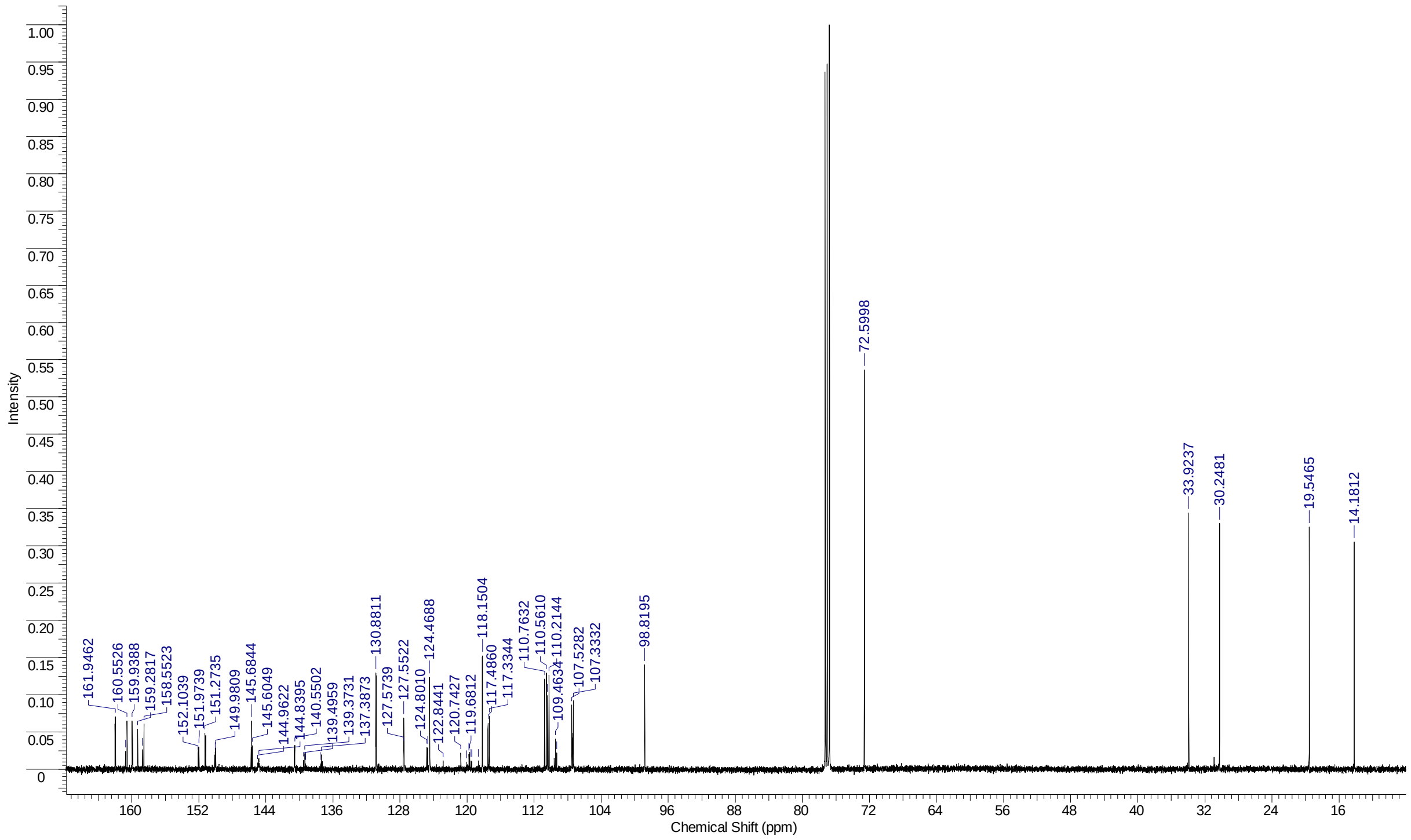

**Suplementarny results**

**Table 1.** Phase transition temperatures (°C) and associated thermal effects (kJ/mol)

| | |
|---|---|
| heating | Cr 71.0 (23.61) $N_F$ 118.0 (0.005) $N_X$ 119.8 (0.12) N 250.6 (1.12) Iso |
| cooling | Iso 237.0 (1.0) N 118.9 (0.085) $N_X$ 114.3 (0.005) $N_F$ 59.0[a] $N_{TBF}$ 55.4 (0.22) $SmC_{TBF}$ 6.6 (7.26) Cr |

[a] from microscopic observations

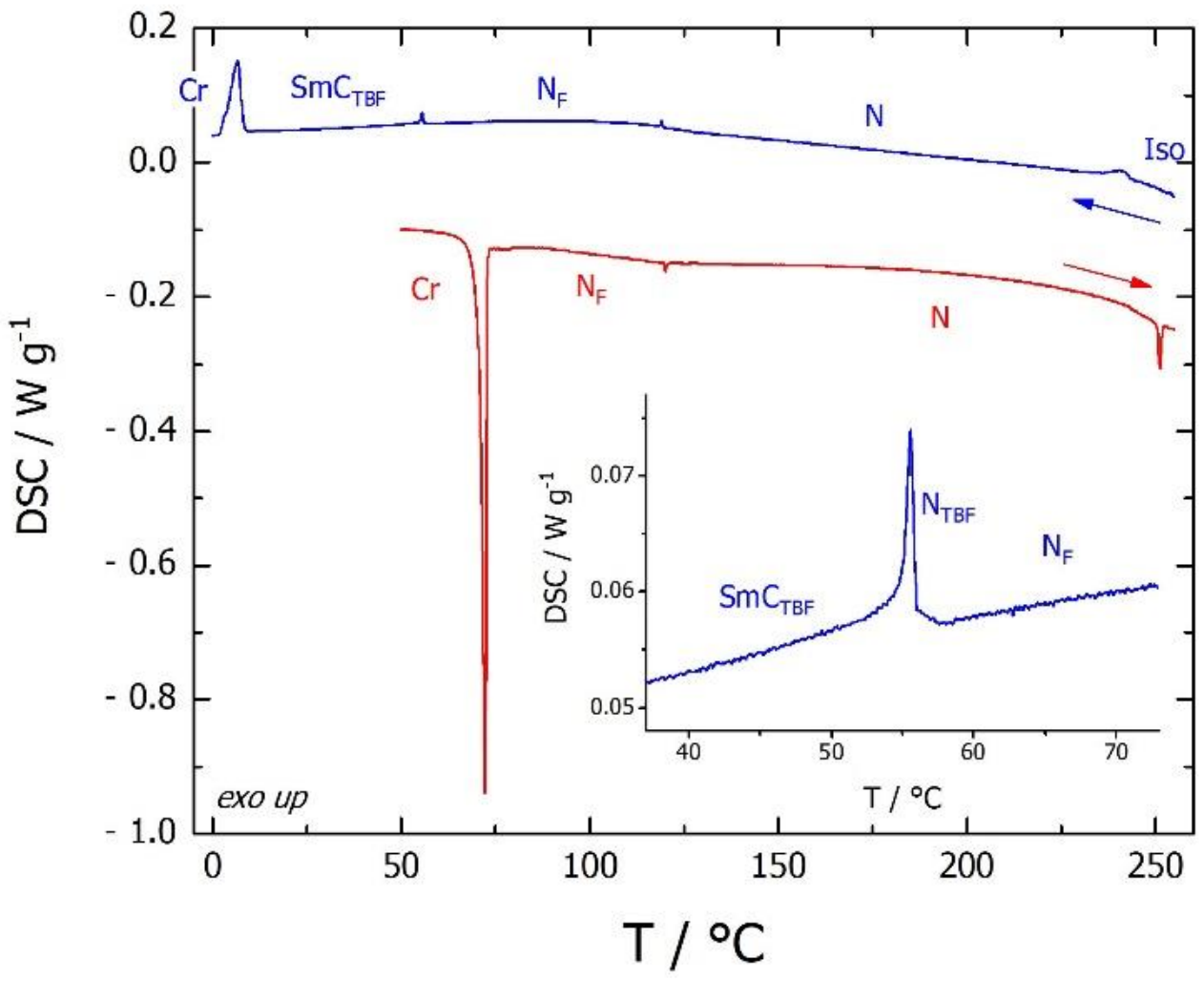

**Figure S1.** DSC thermograms for heating (red line) and cooling (blue line) scans. In the inset: enlarged temperature range near $N_F$-$N_{TBF}$-$SmC_{TBF}$ phase transitions

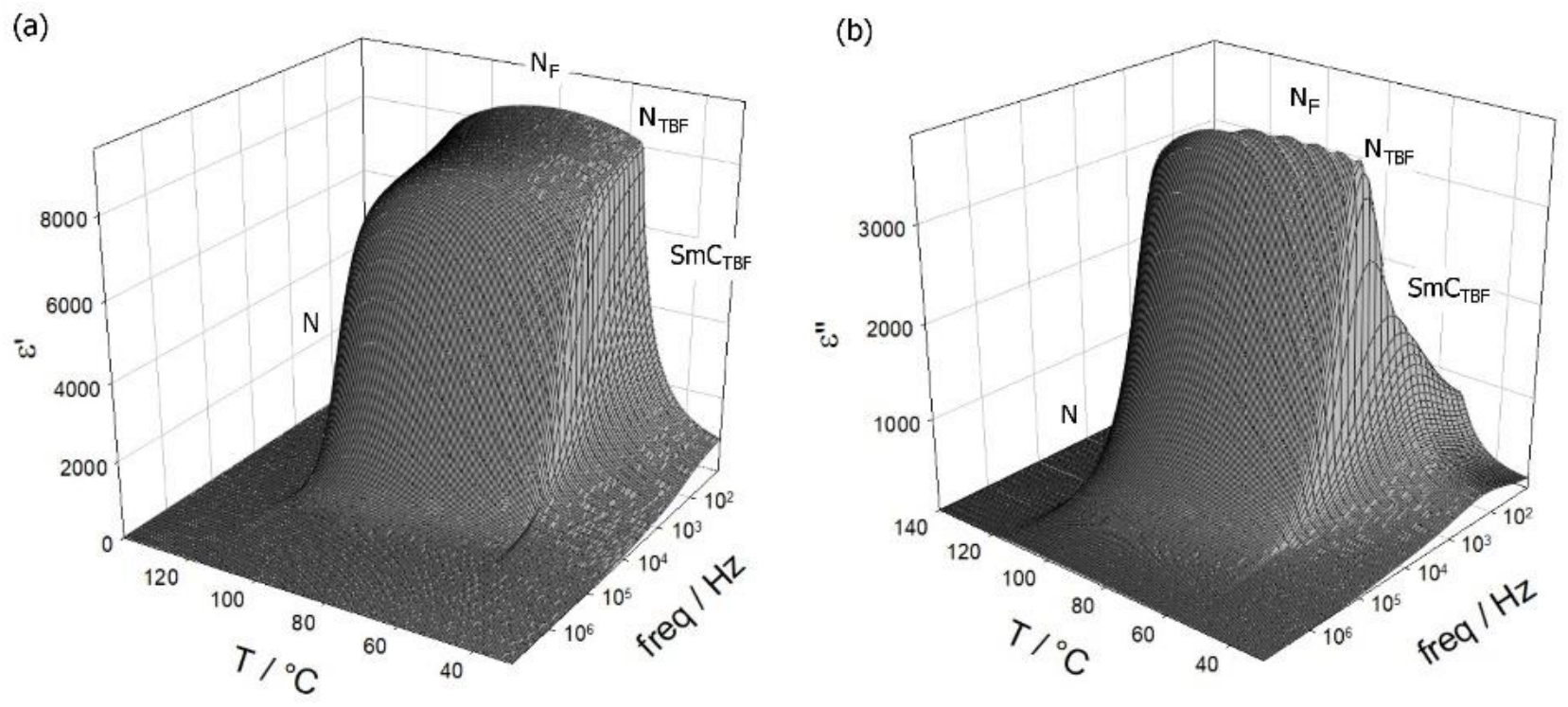

**Figure S2.** Real (a) and imaginary (b) part of dielectric permittivity measured vs. temeprature and frequency.

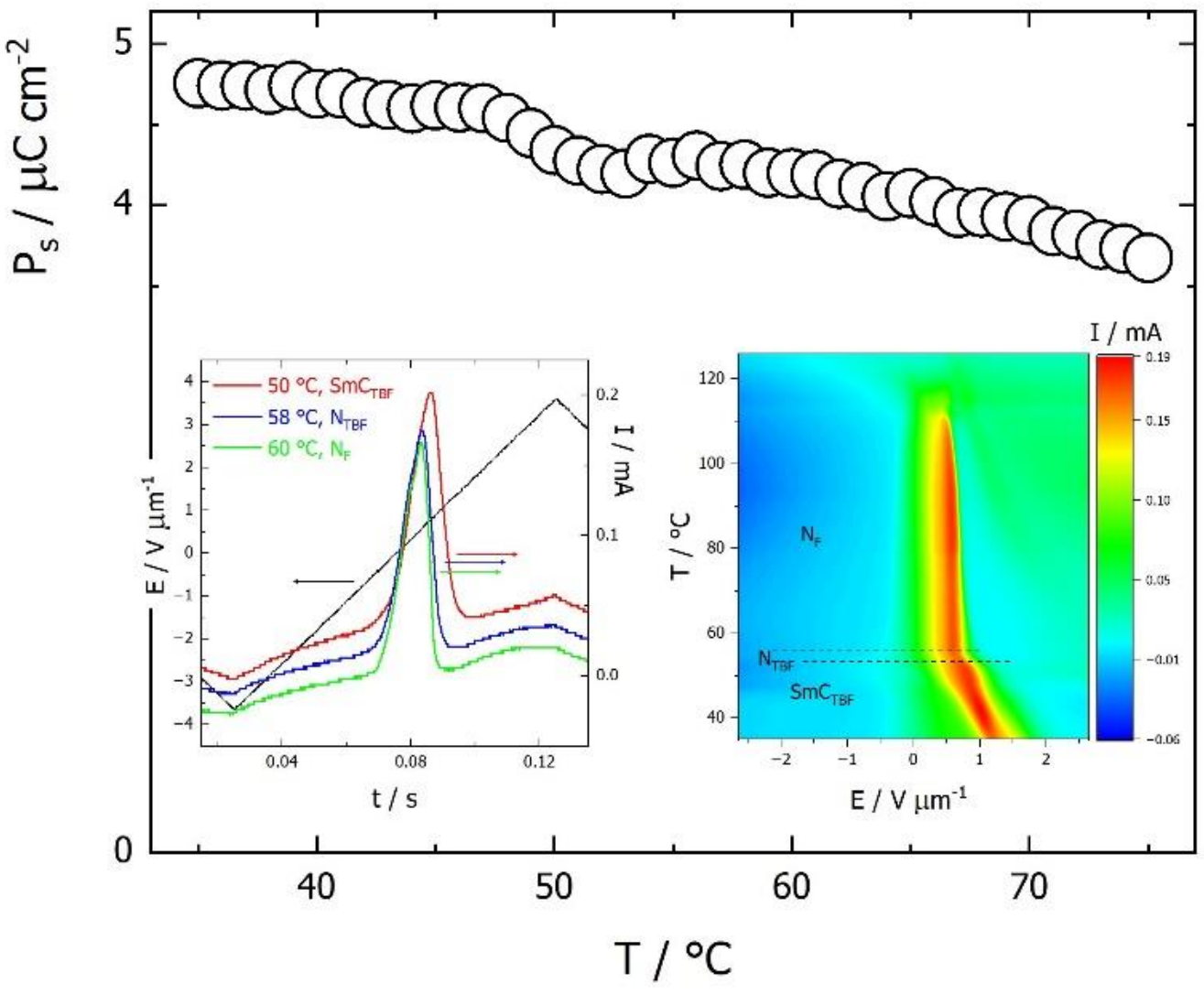

**Figure S3**. Spontanous electric polariztion vs. temperature. In the left inset: switching current in $SmC_{TBF}$, $N_{TBF}$ and $N_F$ phases (red, blue, green lines, respectively) recorded under application of triangular-wave voltage (black line). In the rigth inset: position of the switching currnt peak vs. applied electric field and temperature. In $SmC_{TBF}$ phase the threshold voltage (current peak position) for the switing of polarization considerably increases.

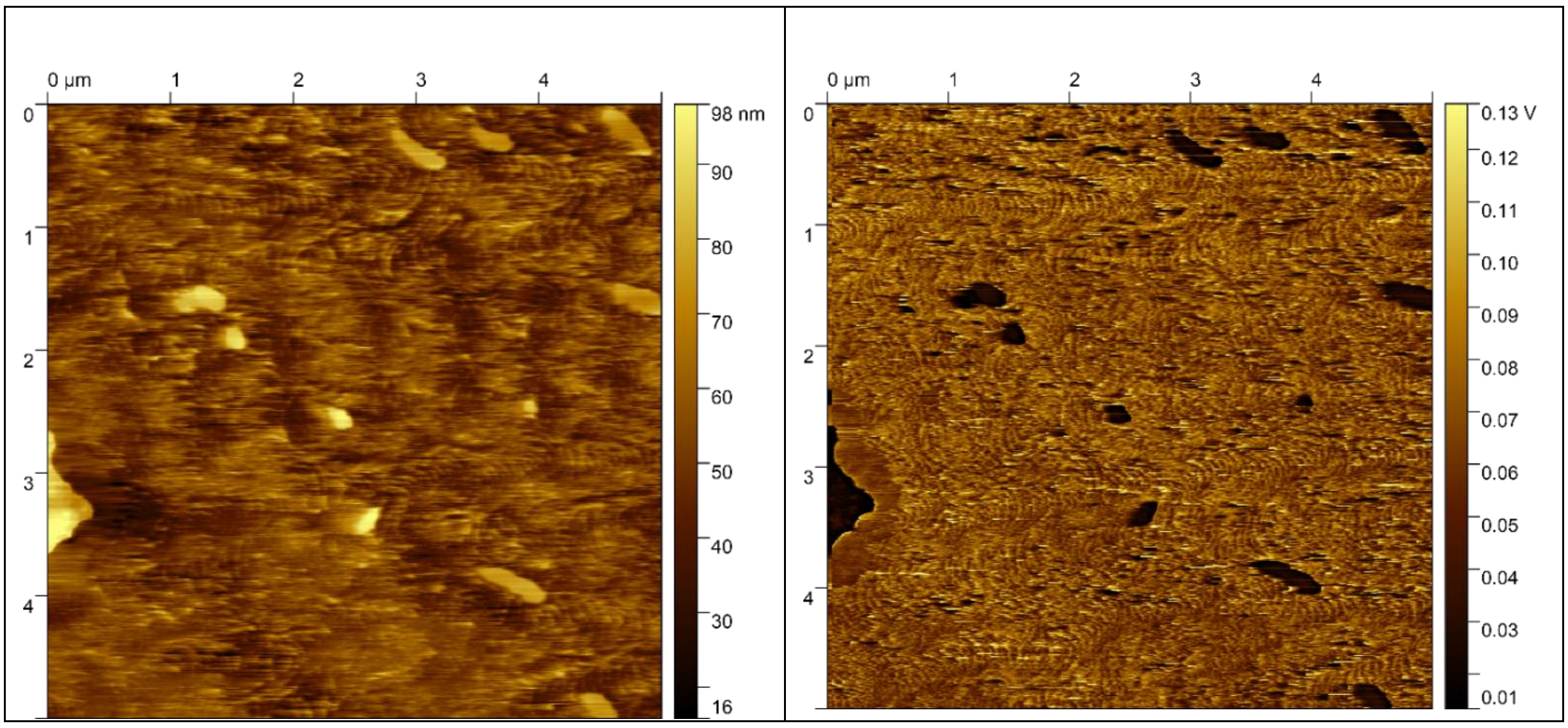

**Figure S4**. AFM image revealing the sub-micrometer structure in the $SmC_{TBF}$ phase: (left) Height and (right) adhesion modes. The weak wavy patterns are most probably the Bouligand arches [25] related to the heliconical structure of the phase.

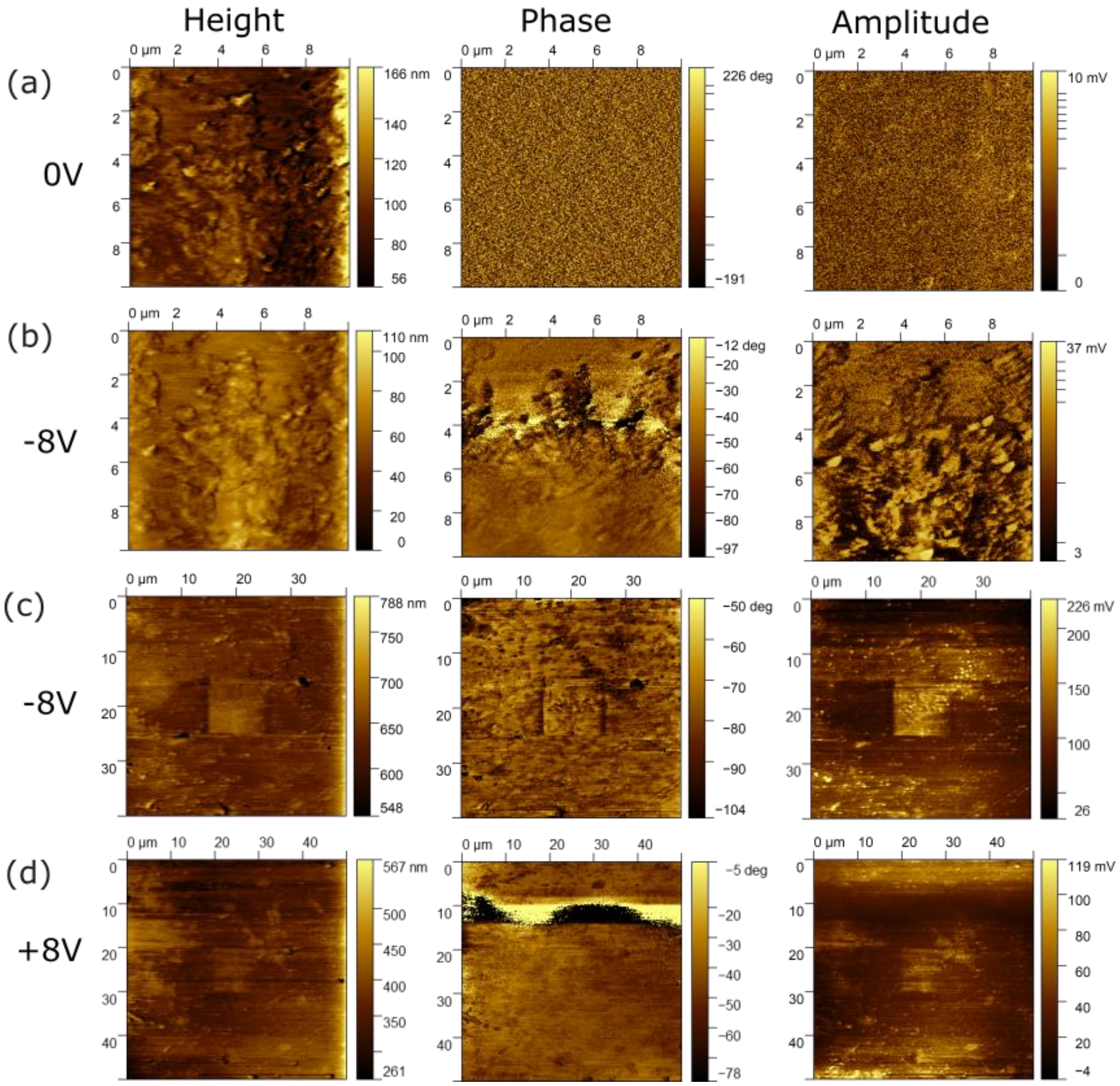

**Figure S5**. (a) 10 x 10 µm sample area measured before application of bias electric field. (b) The same area observed under applied bias voltage (-8V). The measurement under applied bias causes the reorientation of ferroelectric domains. (c) With the -8V bias still applied, measurements were taken over a 40 x 40 µm area. The pre-treated 10 µm square area was clearly visible in the center of the sample. (d) Reversing of the bias voltage to +8V caused the polarization within the 10 µm square domain to be nearly erased. While some remnants of the original domain were still visible in the topography (Height mode), the phase and amplitude were almost completely rewritten.

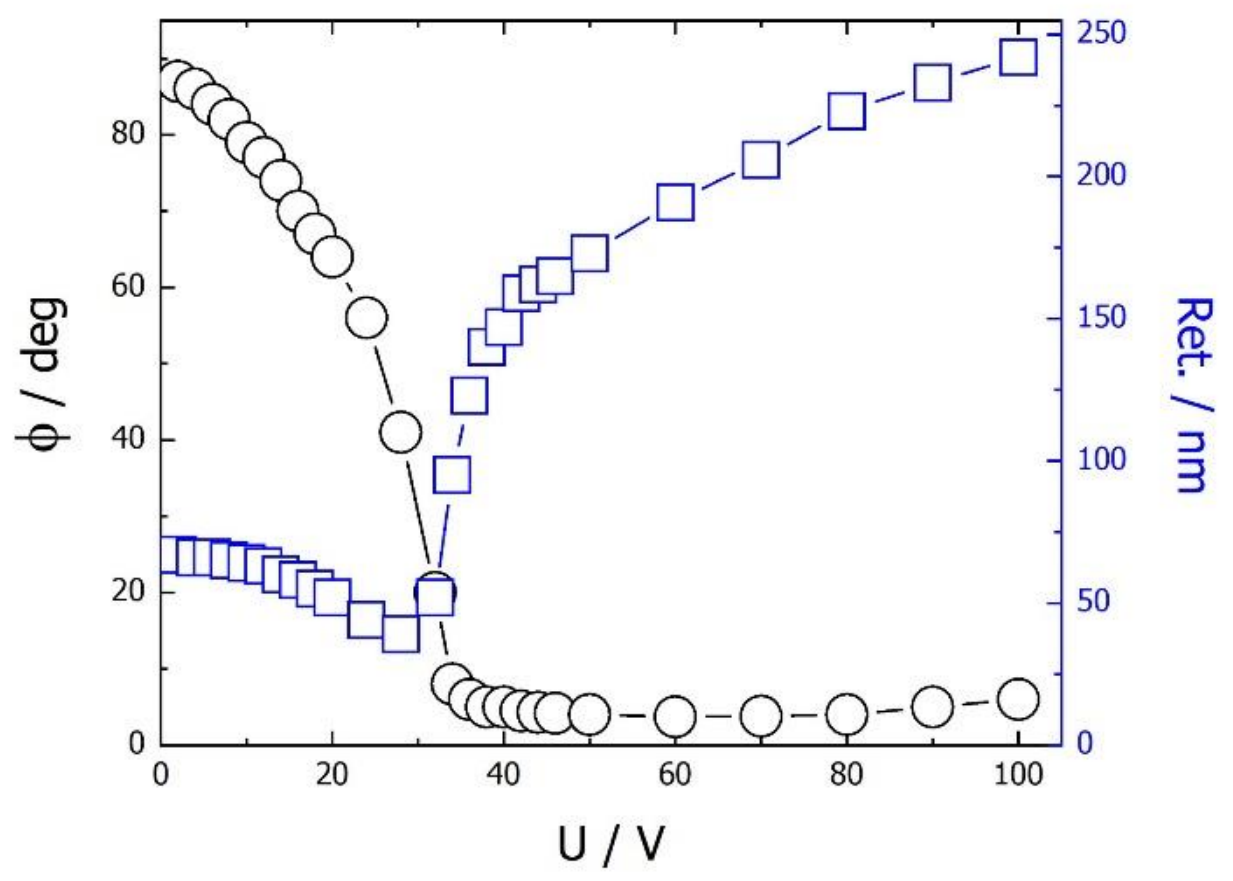

**Figure S6**. Relative orientation of optical axis (□, black circles) and optical retardation (Ret., blue squares) vs. applied voltage measured in $SmC_{TBF}$ phase. A 2.5-□m-thick cell with parallel electrodes on one substrate was used allowing for application of in-plane electric field. The cell surfaces were treated for planar anchoring and were rubbed in direction perpendicular to the electric field.

**Movie S1.** Changes of the laser light diffraction patterns observed on spherical screen in $SmC_{TBF}$ phase (2.5-□m-thick cell with in-plane electrodes, planar anchoring and rubbing direction perpendicular to the electric field) under applied triangular-wave voltage (0.1Hz, 160 $V_{pp}$). Above a certain threshold only diffraction spots positioned along the applied electric field are visible (vertical direction on the movie). When applied electric field is below threshold additional set of diffraction spots appear, along the rubbing direction (horizontal direction on the movie).